\documentclass[11pt]{article}

\usepackage{amsmath}
\usepackage{amsfonts}
\usepackage{amssymb}
\usepackage{graphicx}
\usepackage{supertabular}
\usepackage{setspace}
\usepackage{xcolor}
\usepackage{array}
\usepackage{amsthm}
\usepackage{bm}
\usepackage{latexsym}
\usepackage{mathptmx}
\usepackage{hyperref}
\hypersetup{colorlinks=false}

\RequirePackage{cite}%
\renewcommand{\citeleft}{\bgroup\normalfont[}%
\renewcommand{\citeright}{]\egroup}%

\newcommand{\nin}{\noindent}

\newcommand{\be}{\begin{equation}}
\newcommand{\ee}{\end{equation}}
\newcommand{\ba}{\begin{eqnarray}}
\newcommand{\ea}{\end{eqnarray}}
\newcommand{\bal}{\begin{align}}
\newcommand{\eal}{\end{align}}

\newcommand{\e}{{\rm e}}
\newcommand{\dd}{{\rm d}}

\newcommand{\al}{\alpha}
\newcommand{\la}{\lambda}
\newcommand{\bt}{\beta}
\newcommand{\ka}{\kappa}

\newcommand{\ga}{\gamma}
\newcommand{\ro}{\rho}

\newcommand{\Om}{\Omega}

\newcommand{\bw}{\begin{widetext}}
\newcommand{\ew}{\end{widetext}}

\def\s{\sqrt2}

\def\ps{\phi_s}
\def\rs{\rho_s}
\def\rdm{\ro_{\text{DM}}}
\def\rom{\ro_{\text{OM}}}
\def\P{3-fluid problem}
\def\p{3-fluid problem }

\def\f{scalar field coupled to exponential potential }
\def\S{scalar field problem}
\def\s{scalar field problem }

\begin{document}


\title{{\large\textbf{Numerical solutions to the cosmological 3-fluid problem}}}

\author{Mustapha Azreg-A\"{\i}nou
\\Ba\c{s}kent University, Department of Mathematics, Ba\u{g}l\i ca Campus, Ankara, Turkey}
\date{}

\maketitle

\begin{abstract}
We show that, for the scalar field cosmology with exponential potential, the set of values of the coupling parameter for which the solutions undergo a transient period of acceleration is much larger than the set discussed in the literature. The gradual inclusion of ordinary and dark matters results in an everywhere, but near the origin, smoother and right shifted (along the time axis) acceleration curve. For the 3-fluid problem, the energy density need not exhibit a plateau during the acceleration period. Much excess in the dark matter and/or ordinary matter energy densities would lead the universe to undergo an eternal deceleration expansion. For the 3-fluid problem with a single exponential potential we conclude that the Big Bang Nucleosynthesis constraint is not fulfilled if the universe is to undergo a transient period of acceleration. The 3-fluid model remains a good approximation for the description of large scale structures.

\vspace{3mm}

\nin {\footnotesize\textbf{PACS numbers:} 98.80.-k, 95.35.+d, 95.36.+x, 04.25.D-}

\vspace{-3mm} \nin \line(1,0){430} 
\end{abstract}

\section{The problem} To account for the present period of acceleration of the universe many models have been suggested so far~\cite{1}-\cite{mod}. In this work we are concerned with the models corresponding to spatially flat Friedmann-Roberston-Walker (FRW) universes where the source is (1) a homogeneous \f [representing dark energy (DE)]~\cite{Russo,And} and references therein or (2) a homogeneous \f along with the dark matter (DM), which is modeled by a dust fluid, and a perfect fluid representing ordinary matter (OM)~\cite{3fluid} and references therein, ignoring any explicit coupling of the scalar field to OM or to DM~\cite{SC}, \cite{LA} and~\cite{mod}. This is the so-called \P.

The problem of a \f has been solved in $d$-dimensional by decoupling the system of equations governing the dynamics~\cite{Russo}. If the potential is of the form $V(\phi)=V_0\exp{(-\ka\la\phi)}$ where $\phi$ is the scalar field, the solutions have been classified according to the coupling constant $\la$~\cite{And}. The problem with a piecewise exponential potential in 4-dimensional has also been solved~\cite{And}.

The \p that involves a scalar field (DE), a dust fluid (DM) and a perfect fluid has been treated analytically in~\cite{3fluid}. The method developed in~\cite{3fluid} and references therein that is based on the Noether symmetry assumptions~\cite{ns,GRG} has led, as was shown later in~\cite{And2}, to a flawed solution to the \P, in that the solution derived in~\cite{3fluid} does not satisfy the field equations.

Thus, to our knowledge both problems of deriving analytic solutions or developing new methods to tackle to \p remain open. As we shall see below, even the linear case is not easy to handle. The purpose of this work is to provide numerical solutions to the \p with exponential potential and compare them to the solutions of the \s with similar potential.

To write explicitly the solutions for the normal (not phantom) \s we use a slightly different notation but the same conventions as in~\cite{And}. First, we transform to the new variables and constants (which amounts to take $\ka =\sqrt{3}$):
\begin{equation}\label{c1}
    \la_{\text{new}}=\sqrt{3}\la ,\;\phi_{\text{new}}=\ka\phi/\sqrt{3},\;V_{0\,\text{new}}=\ka^2 V_0/3,\;
    \rho_{\text{new}}=\ka^2 \rho/3,
\end{equation}
where $\rho$ is (any) energy density, then we drop the subscript ``new". With this new notation, the potential takes the form $V(\phi)=V_0\exp{(-\la\phi)}$. For the case of the scalar field alone we use the subscript $s$ to denote any variable. For instance, we write $V_s=V_0\exp{(-\la\ps)}$ and $H_s^2=\rs$ for the \s and $V(\phi)=V_0\exp{(-\la\phi)}$ and $H^2=\rho$ for the \P.

Consider the flat ($k=0$) FRW cosmological model for the universe $\dd s^2=\dd t^2-a_s^2(t)\dd \vec{x}^2$. The equations governing the motion of ($a_s,\ps$) are the FRW equation $H_s^2=\rs$ and the Klein-Gordon one, where $\rs$ is the energy density and $H_s=\dot{a_s}/a_s$ is the Hubble variable ($\dot{x}=\dd x/\dd t$)
\begin{align}
\label{1}&\frac{\dot{a_s}^2}{a_s^2}=\frac{\dot{\ps}^2}{2}+V_0\e^{-\la\ps}\\
\label{2}&\ddot{\ps}+3\frac{\dot{a_s}}{a_s}\dot{\ps}-\la V_0\e^{-\la\ps}=0.
\end{align}
The solutions are called hyperbolic if $\la <3\sqrt{2}$ or trigonometric if $\la >3\sqrt{2}$~\cite{And}. If we let $a_s=\e^{U_s/3}$, $\ps=\sqrt{2}W_s/3$, $\al =\sqrt{18-\la^2}/(3\sqrt{2})$, $\bt =\sqrt{\la^2-18}/(3\sqrt{2})$, and $\ell_{\pm}=6/(6\pm \sqrt{2}\la)$ the hyperbolic and trigonometric solutions are given by, respectively
\begin{align}
\label{3}&U_s=2\tau/\al +\ell_-\ln (1+m\e^{-2 \al  \tau })+\ell_+\ln (1-m\e^{-2\al\tau })\\
\label{4}&W_s=\sqrt{2}\lambda \tau/(3\al) +\ell_-\ln (1+m\e^{-2\al \tau })-\ell_+\ln (1-m\e^{-2\al\tau })\\
\label{15}&U_s=\ell_-\ln [\cos (\bt  \tau )]+\ell_+\ln [\sin (\bt  \tau )]\\
\label{16}&W_s=\ell_-\ln [\cos (\bt  \tau )]-\ell_+\ln [\sin (\bt  \tau )],
\end{align}
where we have omitted additive constants. Here $m$ is a real constant and $\tau$ is the new time coordinate defined by
\begin{equation}\label{5}
    \dot{\tau}=\frac{3}{2}\sqrt{V_0}\e^{-\sqrt{2}\la W_s/6}.
\end{equation}
If $m\neq 0$, then $m$ may be set equal to $\pm 1$ by shifting the values of $\tau$. For all $\la <3\sqrt{2}$, it is straightforward to show that the range of $\tau$ extends from a finite value (usually taken 0) to infinity, and for $\la >3\sqrt{2}$, $0<\tau <\pi/(2\bt)$.

The \p is also governed by the same Klein-Gordon equation~\eqref{2} (with the subscript $s$ removed) and by the FRW equation $H^2=\rho$ with $H=\dot{a}/a$ and $\rho=\rho_{\phi}+\rho_{\text{DM}}+\rho_{\text{OM}}$ is the sum of the energy densities of the three fluids (scalar field, dark and ordinary matters). If the scalar field is normal, its energy density is given by
\begin{equation}\label{6}
   \rho_{\phi}=\frac{\dot{\phi}^2}{2}+V_0\e^{-\la\phi}.
\end{equation}
The DM is treated as a pressureless dust while the OM is a perfect fluid with an equation of state
\begin{equation}\label{7}
    p_{\text{OM}}=(\ga -1)\rho_{\text{OM}}\qquad (1\leq\ga\leq 2).
\end{equation}
Using the three conservation equations (for each fluid $T^{\mu\nu}{}_{;\nu}=0$ where $T^{\mu\nu}$ is the stress-energy tensor) and~\eqref{7}, one arrives at~\cite{3fluid}
\begin{equation}\label{8}
    \rho_{\text{OM}}=Ca^{-3\ga},\quad \rho_{\text{DM}}=E/a^3,
\end{equation}
where $C\geq 0$ and $E\geq 0$ are integration constants. The \p reduces to that of the scalar field if $C=E=0$. The equations of motion take the form\footnote{The second term in Eq. (20) of Ref.~\cite{3fluid} should read $-\varepsilon ka^3\dot{\phi}^2/2$.}
\begin{align}
\label{9}&a\dot{a}^2-\frac{1}{2}a^3\dot{\phi}^2-V_0a^3\e^{-\la\phi}-Ca^{-3(\ga -1)}=E\\
\label{10}&\ddot{\phi}+3\frac{\dot{a}}{a}\dot{\phi}-\la V_0\e^{-\la\phi}=0.
\end{align}
Now we set $a=\e^{U/3}$ and $\phi=\sqrt{2}W/3$ reducing the system to
\begin{align}
\label{11}&9E\e^{-U}=\dot{U}^2-\dot{W}^2-9V_0\e^{-\sqrt{2}\la W/3}-9C\e^{-\ga U}\\
\label{12}&2\ddot{W}+2\dot{U}\dot{W}-3\sqrt{2}\la V_0\e^{-\sqrt{2}\la W/3}=0.
\end{align}
As noticed earlier, if $C=E=0$ the system~\eqref{9} and~\eqref{10} reduces to that of the \S, which is decoupled upon introducing the new variables ($u_s,v_s$) such that $U_s=v_s+u_s$ and $W_s=v_s-u_s$ and using the time $\tau$ defined in~\eqref{5}~\cite{Russo}. It does not seem easy to decouple the system for $C\neq 0$ and $E\neq 0$, the case to which we are interested.

For the purpose of the numerical analysis we are aiming to perform, it is not convenient to introduce a new time coordinate $\tau$ defined by $\dot{\tau}=\frac{3}{2}\sqrt{V_0}\e^{-\sqrt{2}\la W/6}$. Rather, we will use the same time $\tau$ as defined in~\eqref{5}. This will allow us to compare solutions to both problems with the same initial conditions. The system~\eqref{11} and~\eqref{12} takes the form (where $'=\dd /\dd \tau$)
\begin{align}
\label{13}&4E\e^{-U+\sqrt{2} \lambda W_s/3 }=V_0U'^2- V_0W'^2-4V_0\e^{-\sqrt{2}\lambda (W-W_s)/3}-4C\e^{-\gamma U+\sqrt{2}\lambda W_s/3 }\\
\label{14}&6W''-\sqrt{2}\la W_s'W'+6U'W'-4\sqrt{2}\la \e^{-\sqrt{2}\la (W-W_s)/3}.
\end{align}

It is possible by a perturbation approach to decouple the system~\eqref{13} and~\eqref{14} if we assume that the independent constants $C$ and $E$ are small enough to allow for such an approach. In fact, writing $U=U_s+U_CC+U_EE+\cdots$ and $W=W_s+W_CC+W_EE+\cdots$ it is possible to determine the four time functions ($U_C,U_E,W_C,W_E$), however, their expressions are sizable and a numerical analysis is needed anyway. We would prefer applying a numerical approach to the full nonlinear system~\eqref{13} and~\eqref{14} where we can consider large values of $C$ and $E$.

A phase-space analysis of the system~\eqref{9} and~\eqref{10} has been performed in~\cite{phase} and has led to specific solutions (power law inflationary solutions) that provide the late-time attractors or the early-time repellers. However, more other interesting solutions can only be determined numerically as was the case with quintessential cosmological models~\cite{GRG}.

\section{Numerical solutions} 
\begin{figure}[h]
\centering
  \includegraphics[width=0.45\textwidth]{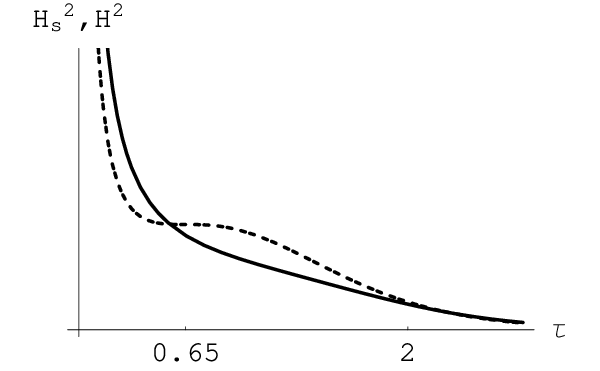} \includegraphics[width=0.45\textwidth]{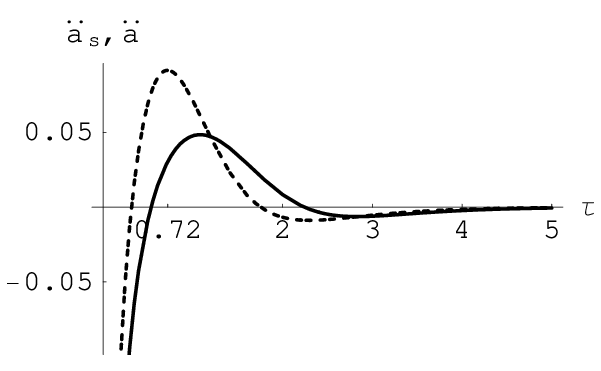} \includegraphics[width=0.45\textwidth]{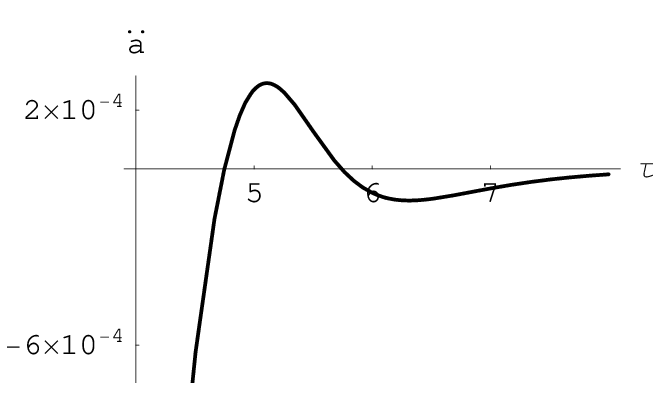}\\
  \caption{\footnotesize{Case $\la =2\sqrt{2}$ and $\ga=1.5$ with $m=1$. (a) (upper left): The energy densities $\rho=H^2$ for $C=E=1$ and $\rs=H_s^2$. $\rs$ has a plateau around $\tau =0.65$. $\rho$ exhibits no plateau where the acceleration occurs. (b) (upper right): The accelerations $\ddot{a}$ for $C=E=1$ and $\ddot{a}_s$. $\ddot{a}_s$ is maximum at $\tau =0.72$. (c) (lower): $\ddot{a}$ for $C=E=100$.}}\label{Fig1}
\end{figure}

For the \S, it was noticed in Ref.~\cite{Russo} that the energy density exhibits a plateau during the transient period of acceleration (TPA). We verify that this statement remains valid for at least\footnote{Because of different conventions, the values of ($\la_R,V_{0R}$) used in~\cite{Russo} are such that $\la =\sqrt{6}\la_R$ and $V_{0R}=6V_0$.} $\sqrt{6}<\la <3\sqrt{2}$ and $\la >3\sqrt{2}$. However, the existence of a plateau for the energy density does not ensure the occurrence of a TPA: We have noticed the existence of solutions, as are the cases for $\la =\sqrt{2}$ and $\la =2/\sqrt{3}$, with no TPA but (twice) the kinetic energy vanishes at the point where the energy density has a plateau and the potential energy is maximum. For instance, for $\la =\sqrt{2}$ a plot of the energy density (with a plateau at $\tau =0.935$), potential and kinetic energies is identical to figure 1 of Ref.~\cite{Russo}. Thus, the clarification provided in~\cite{EG} as to the origin of the TPA lacks support within the scalar field cosmology itself. As we shall see below, in the extended scalar field cosmology, that is the 3-fluid cosmology, the energy density need not exhibit a plateau during the TPA as Fig.~\ref{Fig1} reveals.

For all solutions discussed in this section we choose $V_0=1$ and $\ga =1.5$ and we plot in dashed line (analytic) solutions to the \s and in continuous line (numerical) solutions to the \P. Our initial conditions are $U(\tau_0)=U_s(\tau_0)$, $W(\tau_0)=W_s(\tau_0)$ and $W'(\tau_0)=W_s'(\tau_0)$ with $\tau_0=10^{-10}$.
\begin{figure}[h]
\centering
  \includegraphics[width=0.45\textwidth]{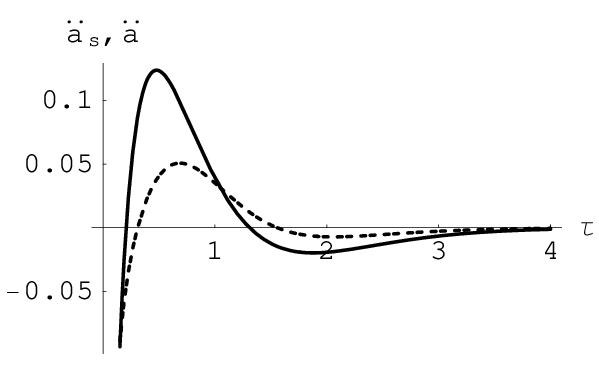}  \includegraphics[width=0.45\textwidth]{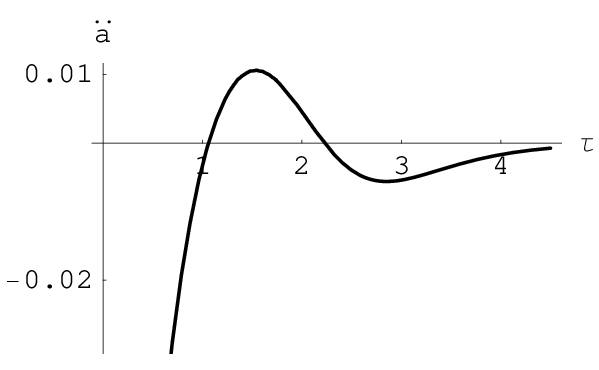} \includegraphics[width=0.45\textwidth]{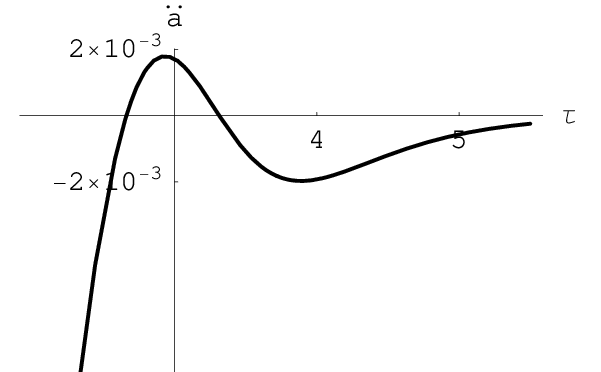}\\
  \caption{\footnotesize{Case $\la =3$ and $\ga=1.5$ with $m=1$. (a) (upper left): $\ddot{a}$ for $C=E=1$ and $\ddot{a}_s$. (b) (upper right): $\ddot{a}$ for $C=E=10$. (c) (lower): $\ddot{a}$ for $C=1,\,E=100$.}}\label{Fig2}
\end{figure}
\begin{figure}[h]
\centering
  \includegraphics[width=0.45\textwidth]{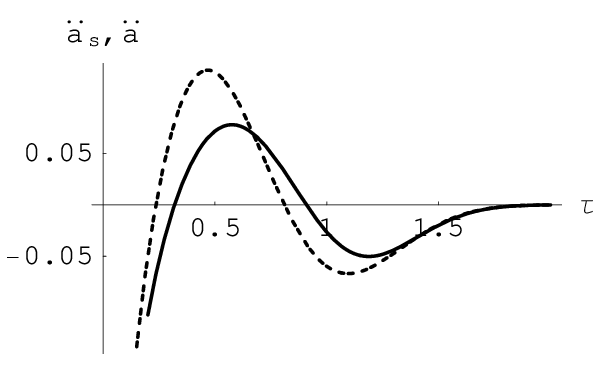}  \includegraphics[width=0.45\textwidth]{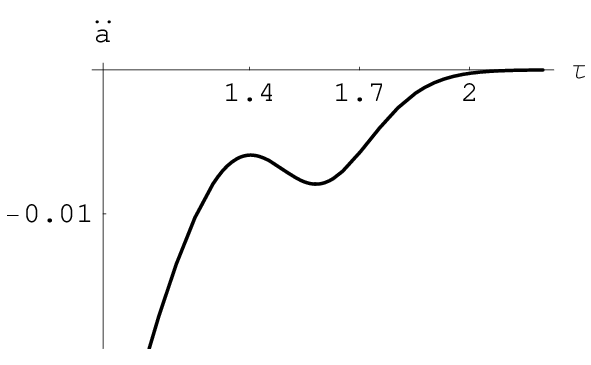} \includegraphics[width=0.45\textwidth]{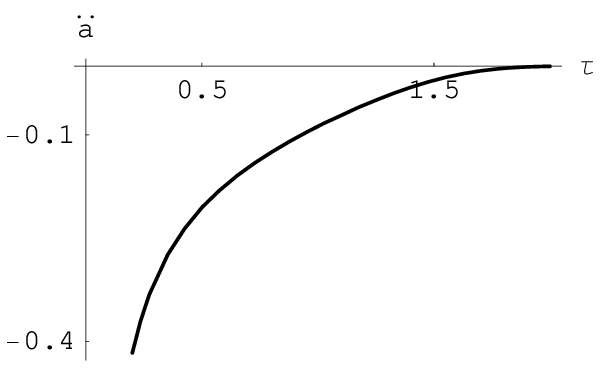}\\
  \caption{\footnotesize{Case $\la =5$ and $\ga=1.5$. (a) (upper left): $\ddot{a}$ for $C=E=1/100$ and $\ddot{a}_s$. (b) (upper right): $\ddot{a}$ for $C=E=1/10$. (c) (lower): $\ddot{a}$ for $C=E=1$.}}\label{Fig3}
\end{figure}

We start by discussing the solution plotted in figure 1 of Ref.~\cite{Russo} for which $\la =2\sqrt{2}$. For $C=E=1$ Fig.~\ref{Fig1} (b) shows plots of $\ddot{a}_s$ and $\ddot{a}$ which both exhibit three phases: deceleration-acceleration-deceleration (DAD). For larger values of the constants $C=E=100$ as in Fig.~\ref{Fig1} (c), the TPA is right-shifted and the acceleration is smoothed. For much larger $C$ and $E$ the acceleration is completely smoothed, except near the origin, and the solution no longer exhibits the DAD phases; rather, it exhibit an eternal decelerated expansion [an example is shown in Fig.~\ref{Fig3} (c) for $\la =5$.] In Fig.~\ref{Fig2} we have similar plots for $\la =3$.


Curiously enough, the trigonometric solutions~\eqref{15} and~\eqref{16} were not discussed deeply in the literature~\cite{Russo,And}. Contrary to a statement made in~\cite{And} which claims that these solutions exhibit only the deceleration-eternal acceleration (DA) expansions, we have checked that for $\la >3\sqrt{2}$ up to (and above) $\la = 100$ these solutions do indeed exhibit the DAD expansions, as the dashed plot in Fig.~\ref{Fig3} (a) depicts. Moreover, the acceleration period coincides with a plateau for the energy density. Numerical solutions to the \p are plotted in Fig.~\ref{Fig3} showing the acceleration for increasing values of the constants. The acceleration phase disappears as the constants increase and the solution ends with an eternal decelerated expansion.

We have checked that the conclusions drawn for the case $\ga =1.5$ extend to the case $\ga =1.9$ and most probably to all $1\leq \ga \leq 2$.

\section{Fine tuning problems and agreement with present observations} Using the dimensionless variables~\cite{p1}
\begin{equation}\label{c2}
    x=\frac{\dot{\phi}}{\sqrt{2}H},\;y=\frac{\sqrt{V}}{H},\;
    z=\frac{\sqrt{\rom}}{H},\;w=\frac{\sqrt{\rdm}}{H},
\end{equation}
which are invariant under transformations~\eqref{c1}, we bring the dynamical equations of the \p to the following system of three linearly independent autonomous differential equations~\cite{phase} [with $N=\ln a(t)$]
\begin{align}
\label{c3}&\frac{\dd x}{\dd N}=\sqrt{\frac{1}{2}} \lambda  y^2-3 x+\frac{3}{2} x [1+x^2-y^2+(\gamma -1) z^2],\\
\label{c4}&\frac{\dd y}{\dd N}=-\sqrt{\frac{1}{2}} \lambda  x y+\frac{3}{2} y [1+x^2-y^2+(\gamma -1) z^2],\\
\label{c5}&\frac{\dd z}{\dd N}=-\frac{3}{2} \gamma  z+\frac{3}{2} z [1+x^2-y^2+(\gamma -1) z^2],\\
&x^2+y^2+z^2+w^2=1.
\end{align}
In terms of the phase-space coordinates, the deceleration parameter $q\equiv -\ddot{a}/(aH^2)$ is such that $2q=1+3[x^2-y^2+(\gamma -1) z^2]$ and the density parameters are given by $\Om_{\phi}= x^2+y^2$, $\Om_{\text{OM}}= z^2$, $\Om_{\text{DM}}= w^2$ and obey the conservation equation $\Om_{\phi}+\Om_{\text{OM}}+\Om_{\text{DM}}=1$.
\begin{figure}[h]
\centering
  \includegraphics[width=0.45\textwidth]{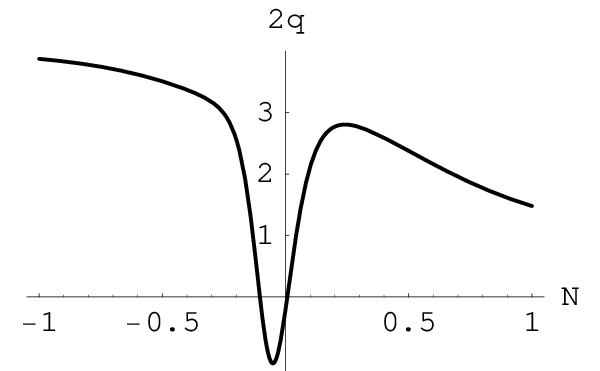}  \includegraphics[width=0.45\textwidth]{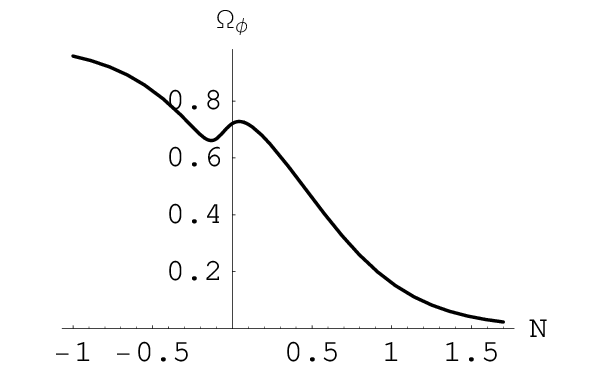} \includegraphics[width=0.45\textwidth]{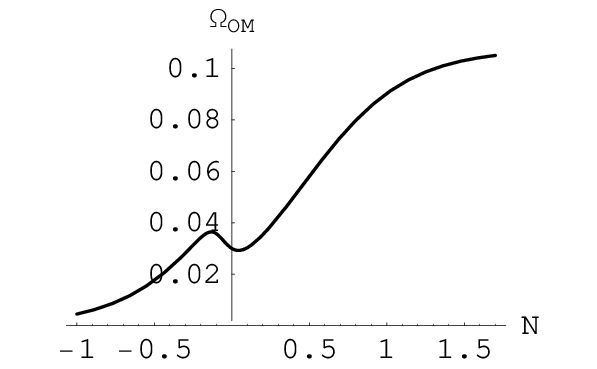} \includegraphics[width=0.45\textwidth]{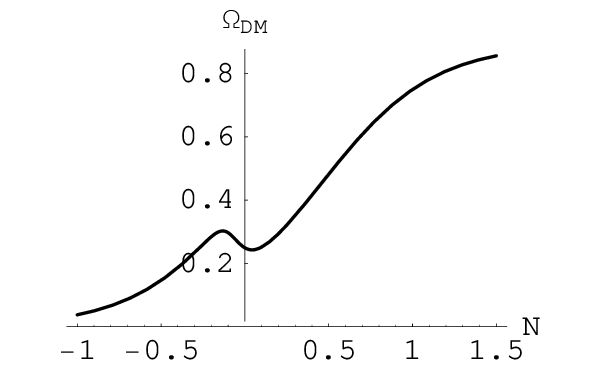}\\
  \caption{\footnotesize{Case $\la =10\sqrt{3}$ and $\ga=1$ with the constraints~\eqref{c6}: End of the TPA. (a) (upper left): Plot of $2q$. The end of the TPA appears near the present age $t_0$ (the plot crosses the $N$ axis at $N=+0.007$ and at another negative value of $N$). (b) (upper right): Plot of $\Om_{\phi}$. At the present age ($N=0$) $\Om_{\phi}$ is nearing its maximum value and then it will decrease to the attractor value $9/\la^2=0.03$. (c) (lower left): Plot of $\Om_{\text{OM}}$. At the present age $\Om_{\text{OM}}$ is nearing its minimum value and then it will increase to the attractor value $0.104301$. (d) (lower right): Plot of $\Om_{\text{DM}}$. At the present age  $\Om_{\text{DM}}$ is nearing its minimum value and then it will increase to the attractor value $0.865699$.}}\label{Fig4}
\end{figure}
\begin{figure}[h]
\centering
  \includegraphics[width=0.45\textwidth]{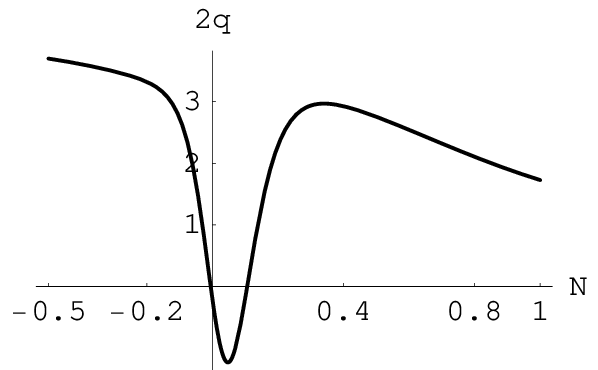}  \includegraphics[width=0.45\textwidth]{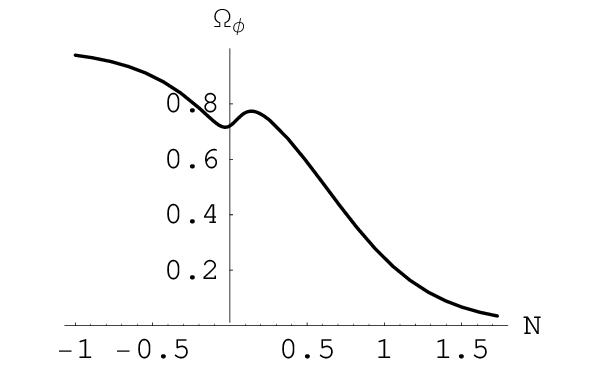} \includegraphics[width=0.45\textwidth]{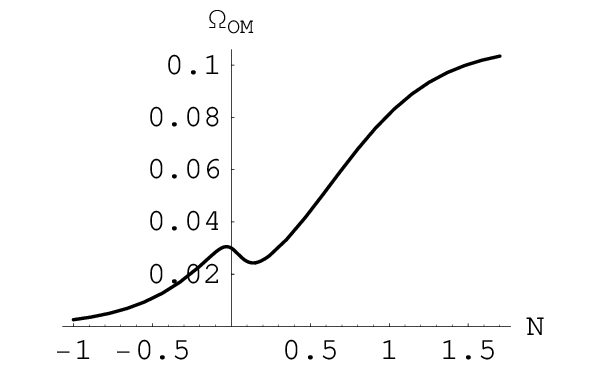} \includegraphics[width=0.45\textwidth]{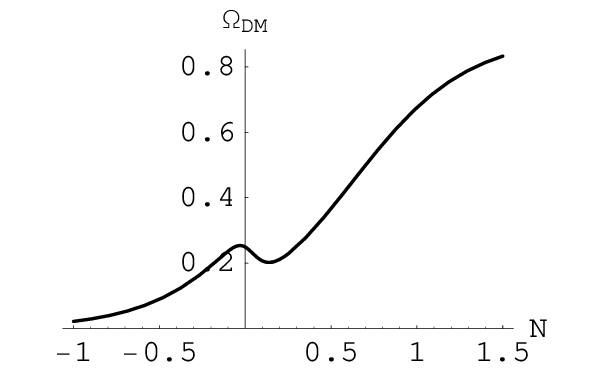}\\
  \caption{\footnotesize{Case $\la =10\sqrt{3}$ and $\ga=1$ with the constraints~\eqref{c6}: Beginning of the TPA. (a) (upper left): Plot of $2q$. The beginning of the TPA appears near the present age $t_0$ (the plot crosses the $N$ axis at $N=-0.005$ and at another positive value of $N$). (b) (upper right): Plot of $\Om_{\phi}$. $\Om_{\phi}$ has just assumed its minimum value near the present age ($N=0$). (c) (lower left): Plot of $\Om_{\text{OM}}$. $\Om_{\text{OM}}$ has just assumed its maximum value near the present age. (d) (lower right): Plot of $\Om_{\text{DM}}$. $\Om_{\text{DM}}$ has just assumed its maximum value near the present age.}}\label{Fig5}
\end{figure}

The modified standard Big Bang Nucleosynthesis (BBN)~\cite{bbn1,bbn2} puts constraints on $\Om_{\phi}$~\cite{bbn1}-~\cite{bbn3} at temperatures $T$ near 1 MeV~\cite{bbn1,bbn2}:
\begin{equation}\label{c6o}
    \Om_{\phi}(1\,\text{MeV})<0.045.
\end{equation}
This corresponds to a constraint on the value of $\la$~\cite{bbn1,bbn2}: $\la>9\sqrt{3}$, thus only the trigonometric solutions~\eqref{15} and~\eqref{16} are concerned [here $\la$ is our new variable defined in~\eqref{c1}]. The other constraints are also observational and correspond to the present density parameters at the present age $t_0$, which are, according to the last and most accurate observations~\cite{obs}, $\Om_{0\,\phi}=0.721\pm 0.015$ and $\Om_0=\Om_{0\,\text{OM}}+\Om_{0\,\text{DM}}= 0.279\pm 0.015$, where $\Om_0$ is the present total matter density parameter. We will work with the values:
\begin{equation}\label{c6}
    \Om_{0\,\phi}=0.721,\;\Om_{0\,\text{OM}}=0.03,\;\Om_{0\,\text{DM}}=0.279-0.03=0.249,
\end{equation}
which are within the limits of observational errors. From now on, we will set $a_0\equiv a(t_0)=1$ [By this last choice the constants in~\eqref{8} are the present densities: $C=\rho_{0\,\text{OM}}$, $E=\rho_{0\,\text{DM}}$].

We use the observational constraints~\eqref{c6} as initial conditions (at the present age, $N=0$) to solve numerically the system~\eqref{c3} to~\eqref{c5}. Figures~\ref{Fig4} and~\ref{Fig5} are plots of twice the deceleration parameter and of the density parameters for $\la =10\sqrt{3}$ and $\ga=1$. By the fine tuning that the constraints~\eqref{c6} define, the universe may be undergoing a TPA which is nearing its end, as Fig.~\ref{Fig4} shows, or it has just started undergoing a TPA, as Fig.~\ref{Fig5} shows. If observations could reveal that $q$ is increasing at the present age or that the total matter density parameter $\Om_0$ is nearing its minimum value as in Fig.~\ref{Fig4}, then the TPA would be nearing its end; otherwise, the TPA would be a recent event in the history of the universe. In both figures, \ref{Fig4} and~\ref{Fig5}, the evolution of the universe approaches that of a late-time attractor~\cite{phase} defined by $\Om_{\phi}=9/\la^2$ (according to the notation of~\cite{phase}, this is the family of attractors $J$).

In figures~\ref{Fig4} and~\ref{Fig5} we have considered the case $\ga=1$ where OM behaves as a dust. Moreover, we have ignored the radiation-dominated era which we could include in a more realistic model as follows. The epoch of matter-radiation equality~\cite{books} corresponds to $N_{\text{eq}}=\ln a_{\text{eq}}=-\ln(24000\Om_0h^2)=-8.15226$ where we take $h=0.72$. Now, since our initial conditions~\eqref{c6} are taken at the present age where radiation is no longer the dominant component, the solutions shown in figures~\ref{Fig4} and~\ref{Fig5} will remain almost unchanged for $N>N_{\text{eq}}$. For $N<N_{\text{eq}}$, the density and deceleration parameters will be (slightly) decreased.

With the constraints~\eqref{c6} imposed, the \p describes well some features of the history of the universe and predicts a TPA at the present age. In our numerical analysis, the present kinetic density $x_0^2$ remains a free parameter constrained by $-\sqrt{0.721}<x_0<\sqrt{0.721}$. However, varying $x_0$ within these limits results mainly in shifting left/right the plots in figures~\ref{Fig4} and~\ref{Fig5}. For instance, we took $x_0=0.4$ and $x_0=-0.4$ in Fig~\ref{Fig4} (a) and Fig~\ref{Fig5} (a), respectively. Thus, the BBN constraint~\eqref{c6o} is satisfied neither in the \s nor in the \p if the model with the single exponential potential is to predict a TPA. Said otherwise, had we chosen the initial conditions so that at the BBN epoch ($N_{\text{BBN}}\sim -20.7$) $\Om_{\phi}(1\,\text{MeV})<0.045$, we would have no TPA at any age following the BBN epoch~\cite{GRG}. This is the result achieved in~\cite{And} where the trigonometric solutions given there do not exhibit any TPA but do satisfy the BBN constraint~\eqref{c6o}. Our trigonometric solutions~\eqref{15} and~\eqref{16} exhibit a TPA because of their extended dominated kination period that prevents the realization of the BBN constraint~\eqref{c6o}.

The conclusion we can draw from figures~\ref{Fig4} (b) and~\ref{Fig5} (b) is that, for the single exponential potential, the kination period, necessary for the occurrence of a TPA, extends beyond the matter dominated era. This has led some authors to modify the potential form and to introduce an interaction term between the dark components~\cite{mod}. As figures 5 and 6 of Ref.~\cite{mod} show, the solutions derived there fulfill the BBN constraint~\eqref{c6o} and undergo a TPA at the present age. For $N\geq 0$, our solutions depicted in figures~\ref{Fig4} and~\ref{Fig5} are in excellent agreement with those depicted in figures 5 and 6 of Ref.~\cite{mod}.

\section{Discussion} Solutions to the \s already predict the DAD expansions for the cosmological evolution of the universe for a wide range of $\la$. Solutions to the \p add new dimensions: The parameters $C$ and $E$ along with $\la$ allow for fine tuning and horizontal and vertical shifting of the acceleration and density curves to fit observational data. Excess in the OM density ($C$ large) or/and the DM density ($E$ large) results in a smooth acceleration curve, except near the origin, exhibiting no DAD phases; rather, the universe would undergo an eternal deceleration expansion.

The BBN constraint $\Om_{\phi}(1\,\text{MeV})<0.045$ is satisfied neither in the \s nor in the \p if the model with the single exponential potential is to predict a TPA.

Solutions with two and probably more TPA's do exist and make the subject of Ref.~\cite{phase}.

In figures~\ref{Fig4} and~\ref{Fig5} we have considered the case $\ga=1$, which is one of the three physical scenarios that the 3-fluid model may fit. As stated in~\cite{phase}, by neglecting all types of interactions, particularly that of visible matter, we restrict the application of the 3-fluid model to beyond the epoch of matter-radiation decoupling, which corresponds to a redshift $\text{z}_{\text{dec}}=1099.9$~\cite{books}. Thus, for $\text{z}<\text{z}_{\text{dec}}$, the model fits well the following three physical scenarios based solely on the value of $\ga$.

\begin{enumerate}
  \item $\ga=1$. The epoch of matter-radiation equality~\cite{books} corresponds to a redshift $\text{z}_{\text{eq}}=24000\Om_0h^2-1=3470.2$ where we take $h=0.72$. With $\text{z}<\text{z}_{\text{dec}}$ it is a good approximation to neglect radiation. The pressureless barotropic fluid represents baryons if the nonrelativistic approximation is valid ($k_{\text{B}}T_b/(m_bc^2)\ll 1$) since in this case the pressure and density obey~\cite{phase}: $p_b=n_bk_{\text{B}}T_b\approx 0$ and $\rho_b=m_bc^2n_b+3n_bk_{\text{B}}T_b/2\approx m_bc^2n_b$ where $n_b$ is the number density and $m_b$ is the rest mass. In this case, there are two pressureless components, the DM and baryons, with a total relative density $\Om_{0}=0.279$ at the present age.
  \item $\ga=4/3$. In this case the components of the universe are regrouped in a way that the barotropic fluid represents radiation, the DM and baryons together make up the pressureless component with a total relative density $\Om_{0}=0.279$, a baryonic density $\Om_{0\,b}=0.04-0.05$ and a DE density $\Om_{0\,\phi}=0.721$~\cite{obs} at the present age.
  \item $\ga<2/3$. This is the case where the universe may undergo (at least) two TPA's and two TPD's. Solutions with two TPA's make the subject of Ref.~\cite{phase}. In this case the barotropic fluid, as the scalar field, has a negative pressure too. Arguing that each component with negative pressure causes a TPA to occur in the history of the universe, we may consider the barotropic fluid as another source of DE. Both sources of DE acting together can be understood as a rough approximation to a more general and elaborate source of DE.

      However, to have a faithful description of the evolution of the universe one should introduce an ordinary matter or baryonic component $\rho_b$ (radiation may be neglected). For a pressureless matter component all that one needs is to add the extra equation
      \begin{equation*}
        u'=3u [x^2-y^2+(\gamma -1) z^2]/2,\qquad (\ga<2/3),
      \end{equation*}
      to the system~\eqref{c3} to~\eqref{c5} with $u=\sqrt{\rho_b}/H$ and $x^2+y^2+z^2+u^2+w^2=1$.

\end{enumerate}

Structure formation is greatly affected by the DE-DM interaction. While there is no observational evidence of the existence of any interaction between the two dark components, some authors, however, arguing that the amounts of DE and DM are comparable at the present age of the universe, have anticipated that and formulated 2- and 3-fluid problems with DE-DM interaction terms~\cite{LA,int,bias,lin,nlin}. The consideration of the DE-DM interaction has led to achieve the following results concerning structure formation. (1) The DE-DM interaction may cause growths in the density perturbations to occur even during the TPA~\cite{bias}, which is not possible without the DE-DM interaction term since, during the TPA, gravity weakens and the perturbation growth ceases. (2) In multiple DM scenarios where the two particle species of DM have opposite couplings to DE~\cite{nlin}, the scalar force does not manifestly affect the large scale structures provided its strength is of order the gravitational force or lower. Even scalar forces twice or three times as large as the gravitational one lead to peculiar features which are identified only in the full nonlinear regime~\cite{nlin}. Thus, up to linear perturbations, the 3-fluid model, we considered here, is a good approximation for the description of large scale structures.


\end{document}